# Zero Magnetic Field Plateau Phase Transition in Higher Chern Number Quantum Anomalous Hall Insulators

Yi-Fan Zhao[1,2], Ruoxi Zhang[1,2], Ling-Jie Zhou[1], Ruobing Mei[1], Zi-Jie Yan[1], Moses H. W. Chan[1], Chao-Xing Liu[1], and Cui-Zu Chang[1]

[1]Department of Physics, The Pennsylvania State University, University Park, PA 16802, USA

[2]These authors contributed equally: Yi-Fan Zhao and Ruoxi Zhang

Corresponding authors: cxc955@psu.edu (C.-Z. C.).

**Abstract:** The plateau-to-plateau transition in quantum Hall effect under high magnetic fields is a celebrated quantum phase transition between two topological states through either sweeping the magnetic field or tuning the carrier density[1-4]. The recent realization of the quantum anomalous Hall (QAH) insulators with tunable Chern numbers introduces the channel degree of freedom to the dissipation-free chiral edge transport and makes the study of the quantum phase transition between two topological states under zero magnetic field possible[5]. Here, we synthesized the magnetic topological insulator (TI)/TI penta-layer heterostructures with different Cr doping concentrations in the middle magnetic TI layers using molecular beam epitaxy (MBE). By performing transport measurements, we found a zero magnetic field quantum phase transition between the $C = 1$ and $C = 2$ QAH states. In tuning the transition, the Hall resistance monotonically decreases from $h/e^2$ to $h/2e^2$, concurrently, the longitudinal resistance exhibits a maximum at the critical point. Our results show that the ratio between the Hall resistance and the longitudinal resistance is greater than 1 at the critical point, which indicates that the original chiral edge channel from the $C = 1$ QAH state coexists with the dissipative bulk conduction channels. Subsequently, **these bulk conduction channels appear to self-organize and form the second chiral edge channel in completing the plateau phase transition. Our study will motivate further investigations of this novel Chern number change-induced quantum phase transition and advance the development of the QAH chiral edge current-based electronic and spintronic devices.**

**Main text:** A rich topic of research in condensed matter physics is to create, manipulate, and understand the quantum phase transition between two different topological states[6,7]. The quantum anomalous Hall (QAH) effect is a prime example of topological states and can be considered as a zero magnetic field manifestation of the quantum Hall effect. The QAH effect, usually realized by time-reversal symmetry breaking in topological nontrivial systems[8-12], possesses a quantized Hall resistance of $h/Ce^2$ with spin-polarized dissipation-free chiral edge channels, where $C$, known as the Chern number, corresponds to the number of chiral edge channels[5,13,14]. Therefore, the QAH effect may have a considerable impact on future electronic and spintronic device applications for ultralow-power consumption. In 2013, the QAH effect was observed in the Cr-doped topological insulator (TI) (Bi, Sb)$_2$Te$_3$ thin films[10]. Two years later in the V-doped TI system, contrary to the prediction from the first-principle calculations[9], a high-precision QAH effect was also demonstrated[15,16]. To date, other than a few exceptions[5,17], most experiments concentrate on $C = 1$ QAH systems[10,14-16,18-33].

Recently, the QAH insulators with tunable Chern number $C$ have been demonstrated in the magnetic TI/TI multilayer heterostructures[5]. A specific magnetic TI/TI penta-layer sample shows either the $C = 1$ QAH effect for the lower magnetic doping concentration or the $C = 2$ QAH effect for the higher magnetic doping concentration. This Chern number change in QAH insulators is induced by the transformation from topological nontrivial to topological trivial in the middle magnetic TI layer[5]. Since the magnetic doping concentration can be varied systematically during the MBE growth process, it is feasible to create a Chern number change-induced quantum phase transition between the $C = 1$ and $C = 2$ QAH states by precisely altering the magnetic doping concentration in the middle magnetic TI layer of the magnetic TI/TI penta-layer samples. The creation of such a Chern number change-induced quantum phase transition under zero magnetic field facilitates the development of chiral edge current based on both high-capacity circuits interconnects and energy-efficient electronic and spintronic devices. Moreover, a more thorough understanding of such a zero magnetic field plateau-to-plateau phase transition enables the inquiry of the fundamental physics of the QAH insulators, such as how the chiral edge transport evolves during the change of the Chern number $C$.

In quantum Hall effect, the plateau-to-plateau transition, which also corresponds to the change of the chiral edge channels (i.e. the change of the Chern number $C$), is usually accessed by either sweeping magnetic field $\mu_0 H$ or tuning the carrier density $n_{2D}$ (Refs. [1-4]). Since the magnetic field $\mu_0 H$ is a prerequisite for the formation of the Landau levels, the Chern number $C$ induced quantum phase transition in quantum Hall effect occurs only under high magnetic fields [1,2]. However, in the QAH effect, the chiral edge transport appears at zero magnetic field[10,15,20,21], the systematic change of the magnetic doping concentration, which leads to the continuous tuning of the Chern number $C$ in QAH insulators, will create a plateau-to-plateau phase transition under zero magnetic field. There are two key questions regarding the plateau-to-plateau phase transition between the $C = 1$ and $C = 2$ QAH insulators: (*i*) What happens to the original chiral edge channel from the initial $C =1$ QAH insulator? (*ii*) How does the chiral edge channel evolve to form the $C =2$ QAH insulator?

In this work, we used molecular beam epitaxy (MBE) to fabricate a series of magnetic TI/TI penta-layer heterostructures, specifically 3 quintuple layers (QL) (Bi, Sb)$_{1.73}$Cr$_{0.27}$Te$_3$/4QL (Bi,

Sb)$_2$Te$_3$/3 QL (Bi, Sb)$_{2-x}$Cr$_x$Te$_3$/4 QL (Bi, Sb)$_2$Te$_3$ /3 QL (Bi, Sb)$_{1.73}$Cr$_{0.27}$Te$_3$), by systematically varying the Cr doping concentration $x$ in the middle magnetic TI layer. We performed transport measurements and observed a plateau-to-plateau phase transition between the $C = 1$ to $C = 2$ QAH insulators under zero magnetic field (Fig. 1a). We found that the ratio between the Hall resistance and the longitudinal resistance under zero magnetic field is greater than 1, suggesting the sample still resides in a nonperfect QAH state at the critical point of the plateau-to-plateau phase transition[5]. This observation implies that the original chiral edge channel from the $C = 1$ QAH insulator coexists with the bulk conducting channel during the Chern number change-induced quantum phase transition.

All magnetic TI/TI penta-layer heterostructures used in this work were grown on the heat-treated 0.5 mm thick SrTiO$_3$(111) substrates in an MBE chamber (Omicron Lab 10) with a base pressure ~2 × 10$^{-10}$ mbar. To make all samples consistent over the entire quantum phase transition regime, we maintained the Cr doping concentration in the top and bottom magnetic TI layers at $x$ = 0.27 and systematically changed the Cr doping concentration $x$ in the middle magnetic TI layer from $x$ = 0.08 to $x$ = 0.26 (Fig. 1a). The Bi/Sb ratio in each layer was optimized to tune the chemical potential of the sample near the charge neutral point. The electrical transport measurements were carried out in a Physical Property Measurements System (Quantum Design DynaCool, 2 K, 9 T) and a dilution refrigerator (Leiden Cryogenics, 10 mK, 9 T) with the magnetic field applied perpendicular to the film plane. Six terminal mechanically defined Hall bars were used for electrical transport measurements. More details about the MBE growth of the samples and electrical transport measurements can be found in the Methods section.

We first performed transport measurements on these MBE-grown magnetic TI/TI penta-layer heterostructures with different $x$ at $T$ = 25 mK and the charge neutral point $V_g = V_g^0$. The $V_g^0$ here is determined when the zero magnetic field Hall resistance (labeled as $\rho_{yx}(0)$) is maximized. The $C=1$ QAH effect is realized in a sample with $x = 0.08$ where the Hall resistance $\rho_{yx}$ is found to be 0.986 $h/e^2$, and the longitudinal resistance $\rho_{xx}$ is 0.0009 $h/e^2$ (~ 23 Ω) under zero magnetic field (Fig. 1b). For the $x = 0.15$ sample, $\rho_{yx} \sim 0.977\ h/e^2$ and $\rho_{xx} \sim 0.0236\ h/e^2$ (~ 600Ω) under zero magnetic field (Fig. 1c), slightly deviating from the perfect $C=1$ QAH state. By steadily increasing the Cr doping level $x$, $\rho_{yx}$ further deviates from the quantized values of $\sim h/e^2$ and finally saturates at the half-quantized value of $\sim h/2e^2$, corresponding to the quantum phase transition between the $C = 1$ to $C = 2$ QAH insulators. $\rho_{yx}$ values are found to be 0.797 $h/e^2$, 0.655 $h/e^2$, 0.615 $h/e^2$, and 0.494 $h/e^2$ (i.e., the $C = 2$ QAH effect) for the $x = 0.18$, 0.19, 0.20, and 0.26 samples, respectively, under zero magnetic field. The corresponding values of $\rho_{xx}$ are 0.223 $h/e^2$, 0.456 $h/e^2$, 0.320 $h/e^2$, and 0.0015 $h/e^2$ (~ 40Ω), respectively (Figs. 1d to 1g). The large $\rho_{xx}$ in the $x = 0.18$, 0.19, and 0.20 samples implies that the bulk conduction channels are present in these three samples in the transition region. Compared to the well-quantized $\rho_{yx}$ in both the $x = 0.08$ and 0.26 samples (i.e., the $C = 1$ and $C = 2$ QAH insulators), both $\rho_{yx}$ and $\rho_{xx}$ of the other four samples show kink features after the magnetic field $\mu_0 H$ crosses the zero magnetic field, which become more pronounced when the QAH samples are away from the quantized regimes. This kink feature is likely related to temperature rise in the sample induced by magnetic field polarity reversal during the magnetic field sweep[5,31,34]. Moreover, we noted that a topological Hall-like hump feature also appears during the magnetization reversal process in the samples that are located in the quantum phase transition regime. This observation suggests that the chiral magnetic domain walls and/or other chiral spin textures might be present in these magnetic TI heterostructure samples[32,35].

To further understand the plateau phase transition between the $C = 1$ and $C = 2$ QAH insulators, we measured magnetic field $\mu_0 H$ dependence of $\rho_{yx}$ and $\rho_{xx}$ at different gate $V_g$ values and plotted the $V_g$ dependence of $\rho_{yx}(0)$ and zero magnetic field $\rho_{xx}$ (labeled as $\rho_{xx}(0)$) in Fig. 2. For the $x =$ 0.08 and $x = 0.26$ samples, $\rho_{yx}(0)$ exhibits a distinct plateau centered at $V_g = V_g^0$ with the quantized values of $h/e^2$ and $h/2e^2$, respectively. The corresponding $\rho_{xx}(0)$ shows a wide zero resistance plateau, validating the perfect $C = 1$ and $C = 2$ QAH insulator states (Figs. 2a and 2f). For the $x = $ 0.15, 0.18, 0.19, and 0.20 samples, $\rho_{yx}(0)$ shows a maximum at $V_g = V_g^0$ (Figs. 2b to 2e). The corresponding $\rho_{xx}(0)$ shows a dip feature at $V_g = V_g^0$ for the $x = 0.15$ and 0.18 samples. For the $x = $ 0.19 sample, $\rho_{xx}(0)$ also shows a dip feature centering slightly away from $V_g = V_g^0$. However, this dip feature in $\rho_{xx}(0)$ is not seen in the $x = 0.20$ sample. For the three samples located in the quantum phase transition regime, the $\rho_{yx}(0)/\rho_{xx}(0)$ ratios are 3.57, 1.44, and 1.92, corresponding to the Hall angles of 74.37°, 55.15°, and 62.51° for the 0.18, 0.19, and 0.20 samples, respectively. According to the criterion for the emergence of the QAH state, which is defined as $\rho_{yx}(0)/\rho_{xx}(0) \geq 1$, the original chiral edge channel from the $C = 1$ QAH insulator persists and never disappears in the phase transition region between the $C = 1$ and $C = 2$ QAH insulators. We noted that although both the $x = 0.19$ and 0.20 samples have the chiral edge transport, i.e., in the nonperfect QAH regime, the misalignment between the $\rho_{yx}$ maximum and the $\rho_{xx}$ dip in the $x = 0.19$ sample and the absence of the $\rho_{xx}$ dip in the $x = 0.20$ sample are unusual. We speculated that these two behaviors might be attributed to the coexistence of the chiral edge transport and the bulk conducting transport during the Chern number change-induced quantum phase transition. More studies are required to clarify their underlying physics.

Figure 3a shows the dependence of $\rho_{yx}(0)$ and $\rho_{xx}(0)$ on the Cr doping concentration $x$ in the middle magnetic TI layer. Eleven samples were measured at $T = 25$ mK and $V_g = V_g^0$. These eleven samples of different Cr concentrations $x$ show a systematic and smooth evolution from the $C = 1$

QAH state through a transition region and end in the $C = 2$ QAH state. If we define $\rho_{xx}(0) < 0.02$ $h/e^2$ as the criterion for the quantized states, then the $C = 1$ and $C = 2$ QAH states are found for $x < 0.14$ and $x > 0.26$ samples, respectively. For $0.14 \leq x \leq 0.26$, $\rho_{yx}(0)$ smoothly changes from the quantized $h/e^2$ plateau to the adjacent quantized $h/2e^2$ plateau. Concurrently, $\rho_{xx}(0)$ exhibits a peak at $x \sim 0.19$. These observations validate the $x$ change-induced quantum phase transition between the $C = 1$ and $C = 2$ QAH insulators. Note that the plateau-to-plateau quantum phase transition observed here is analogous to that found in the quantum Hall effect but realized under zero magnetic field.

During the plateau phase transition regime, $\rho_{yx}(0)$ is always greater than $\rho_{xx}(0)$, i.e. $\rho_{yx}(0)/\rho_{xx}(0) > 1$ (Fig. 3b). This indicates that throughout the transition, the original chiral edge channel persists and coexists with the bulk conducting channels introduced by the topological nontrivial to the topological trivial quantum phase transition of the middle magnetic TI layer[5,36,37]. In the following, we presented a phenomenological picture to understand this Cr doping concentration $x$ induced Chern number quantum phase transition. In magnetic TI/TI penta-layer heterostructures, the total Hall conductance $\sigma_{xy}$ prominently comes from the gapped Dirac bands located at the interfaces between the magnetic TI layers and the undoped TI layers. Each nontrivial interface state contributes $e^2/2h$. For the $x < 0.14$ samples, the middle magnetic TI is still located in the topological nontrivial regime and shares the same topology with the undoped TI layer, so there are only two nontrivial interfaces states located at the two outer magnetic TI/TI interfaces in the penta-layer heterostructure samples, and thus the total Hall conductance $\sigma_{xy} \sim e^2/h$, corresponding to the $C = 1$ QAH state. For the $x > 0.26$ samples, the middle magnetic TI becomes a trivial insulator and thus two more nontrivial interface states appear at the two inner interfaces between the undoped TI layers and the middle magnetic TI layers. The total Hall conductance $\sigma_{xy}$

~ $2e^2/h$, corresponding to the $C = 2$ QAH state. For $0.14 \leq x \leq 0.26$, increasing $x$ makes the middle magnetic TI layer undergo a phase transition from being topological nontrivial to topological trivial due to the reduction of the spin-orbital coupling of the middle magnetic TI layer[5,36,37]. This topological phase transition is responsible for the occurrence of the Chern number $C$ change-induced quantum phase transition observed in QAH penta-layer structures. To support this phenomenological picture, we performed numerical simulations on the band dispersions and wavefunction distributions for the magnetic TI/TI penta-layer heterostructures (Figs. S15 and S16). Our calculations indeed show a topological phase transition between the $C = 1$ to $C = 2$ QAH states by solely varying the parameters that account for the exchange coupling and spin-orbit coupling strength in the middle magnetic TI layer (Fig. S15). The wavefunction distributions of the lowest-energy conduction bands along the $z$-direction also reveal a variation from the bulk state formed with a single peak to the interface states formed with double peaks across the phase transition (Fig. S16). This validates our phenomenological picture for the origin of the zero magnetic field plateau phase transition in magnetic TI penta-layer samples.

Finally, we discussed the change of the critical temperatures of the QAH states within the Chern number change-induced phase transition regime. As noted above, the critical temperature $T_c$ of the QAH state is defined as the temperature at which $\rho_{yx}(0)/\rho_{xx}(0) = 1$, i.e. the crossing point of the temperature dependence of $\rho_{yx}(0)$ and $\rho_{xx}(0)$ curves (Fig.4). The $T_c$ values of the QAH states for the six samples are shown in Figs. 1 and 2 are ~6.8 K, ~2.2 K, ~1.2 K, ~0.2 K, ~1.8 K, and ~9.2 K for the $x$ = 0.08, 0.15, 0.18, 0.19, 0.20, and 0.26 samples, respectively. The decrease of $\rho_{xx}(0)$ with decreasing temperature in the $x$ = 0.18, 0.19, and 0.20 further confirms that these three samples are still located in the nonperfect QAH regime[15,32]. During the tuning of the transition from the $C = 1$ to $C = 2$ QAH states, the introduction of the bulk conducting channels in the middle

TI layers obscure the original chiral edge transport from the $C = 1$ QAH state. Concomitantly, when the middle magnetic TI layer becomes a trivial insulator, the second chiral edge channel gradually emerges. This transformation completes with the sample becoming the $C = 2$ QAH insulator with two chiral edge channels. The relatively low critical temperature in the $x = 0.19$ sample at the transition critical point also supports that the original chiral edge channel is indeed affected by the appearance of the bulk conducting channels.

To summarize, we fabricated magnetic TI/TI penta-layer samples with different Cr doping concentrations in the middle magnetic TI layer and realized the zero magnetic field quantum phase transition between the $C = 1$ to $C = 2$ QAH insulators. We found that the Hall resistance exhibits a plateau-to-plateau transition and the longitudinal resistance shows a maximum at the transition critical point. We also demonstrated that through the phase transition, the original chiral edge channel from the $C = 1$ QAH insulator persists and never disappears, concomitantly, the coexisting bulk conduction channels self-organize into one additional chiral edge channel upon the completion of the transition. Our observation of the coexistence of the chiral edge states and the dissipative bulk conducting channels paves the way for future quantitative studies on the scaling behaviors of these novel quantum phase transitions in QAH insulators and also lays down the necessary conditions and limitations on how the QAH insulators can be utilized in next-generation electronic and spintronic devices with low power consumption.

**Methods**

**MBE growth**

The magnetic TI/TI penta-layer heterostructures used in this work were grown on heat-treated 0.5 mm thick $SrTiO_3(111)$ substrates in a commercial MBE system (Omicron Lab 10) with a base vacuum better than $2 \times 10^{-10}$ mbar. Before the MBE growth of the magnetic TI/TI penta-layer heterostructures, the heat-treated SrTiO$_3$(111) substrates were first outgassed at ~ 600 °C for an hour. High purity Cr(99.999%), Bi(99.9999%), Sb(99.9999%), and Te(99.9999%) were evaporated from Knudsen effusion cells. The Cr doping concentration $x$ was precisely controlled by the evaporation temperature of the Cr cell, and the value of $x$ was calibrated by the determination of the Curie temperatures of the 6 QL Sb$_{2-x}$Cr$_x$Te$_3$ samples in a prior report (see details in Supplementary Information)[15]. During the growth, the substrates were maintained at ~230 °C. The flux ratio of Te per (Cr + Bi + Sb) was set to greater than 10 to prevent Te deficiency in the films. The Bi/Sb ratio in each layer was optimized to tune the chemical potential of the entire penta-layer heterostructure near the charge neutral point. The growths of magnetic TI and TI films are uninterruptedly grown. The growth rates of both films were ~ 0.2 QL/minutes.

**Electrical transport measurements**

All magnetic TI/TI penta-layer heterostructures grown on the 2 mm × 10 mm insulating SrTiO$_3$(111) substrates for electrical transport measurements were scratched into a Hall bar geometry using a computer-controlled probe station. The effective area of the Hall bar is ~1 mm × 0.5 mm. The electrical ohmic contacts were made by pressing indium dots on the films. The bottom gate was prepared by flattening the indium dots on the back side of the SrTiO$_3$(111) substrates. Transport measurements were conducted using a Physical Property Measurement System (Quantum Design DynaCool, 2 K, 9 T) for $T \geq 2$K and a Leiden Cryogenics dilution refrigerator (10 mK, 9 T) for $T < 2$ K. The excitation currents are 1 μA and 1 nA for the PPMS and the dilution measurements, respectively. All magneto-transport results shown in this paper were symmetrized or anti-symmetrized as a function of the magnetic field to eliminate the influence of the electrode misalignment. More transport results can be found in the Supplementary Information.

**Acknowledgments:** We thank Yongtao Cui, Nitin Samarth, Di Xiao, Xiaodong Xu, Jiaqiang Yan

for helpful discussions. The work is primarily supported by the DOE grant (DE-SC0019064), including the MBE growth, the dilution temperature electrical transport measurements, and the theoretical support. The PPMS measurements were partially supported by an ARO Young Investigator Program Award (W911NF1810198), an NSF-CAREER award (DMR-1847811), and the Gordon and Betty Moore Foundation's EPiQS Initiative (GBMF9063 to C.-Z. C.). Part of the measurements at dilution-refrigerator temperature is supported by NSF grant DMR-1707340.

**Author contributions:** C.-Z. C. conceived and designed the experiment. Y.-F. Z., L.-J. Z., and Z.-J. Y. grew all magnetic TI/TI penta-layer heterostructures and carried out the PPMS transport measurements with the help of C.-Z. C. R. Z., Y.-F. Z., and L.-J. Z. performed the dilution measurements with the help of M. H. W. C. and C.-Z. C.. R. M and C.-X. L. provided theoretical support. Y. -F. Z. and C. -Z. C. analyzed the data and wrote the manuscript with inputs from all authors.

**Competing interests:** The authors declare no competing interests.

**Data availability:** The datasets generated during and/or analyzed during this study are available from the corresponding author upon reasonable request.

**Code availability:** The codes used in theoretical simulations and calculations are available from the corresponding author upon reasonable request.

**Figures and figure captions:**

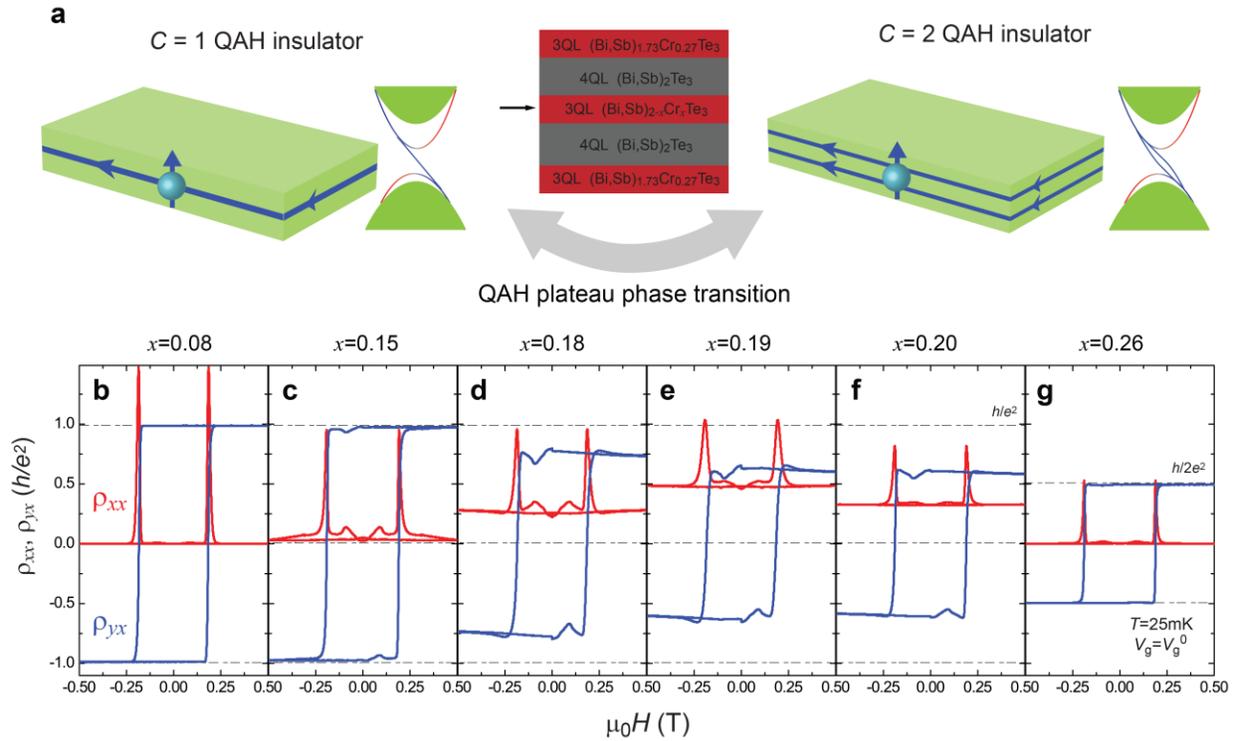

**Figure 1| QAH plateau phase transition under zero magnetic field. a,** Left (right): Schematic and the corresponding electronic band structure of the $C = 1$ ($C = 2$) QAH insulator. Middle: Schematic of the magnetic TI/TI penta-layer heterostructure, specifically 3QL (Bi, Sb)$_{1.73}$Cr$_{0.27}$Te$_3$/4QL (Bi, Sb)$_2$Te$_3$/3QL (Bi, Sb)$_{2-x}$Cr$_x$Te$_3$/3QL (Bi, Sb)$_{1.73}$Cr$_{0.27}$Te$_3$/4QL (Bi, Sb)$_2$Te$_3$. Varying $x$ in the middle magnetic TI layer induces a phase transition between $C=1$ and $C=2$ QAH insulators. **b-g,** $\mu_0 H$ dependence of $\rho_{xx}$ (red) and $\rho_{yx}$ (blue) measured at $V_g = V_g^0$ and $T = 25$ mK. $V_g^0$ values are +4 V, +20 V, -1 V, -23 V, +14 V, and +20 V for the $x$=0.08, $x$=0.15, $x$=0.18, $x$=0.19, $x$=0.20, and $x$=0.26, respectively.

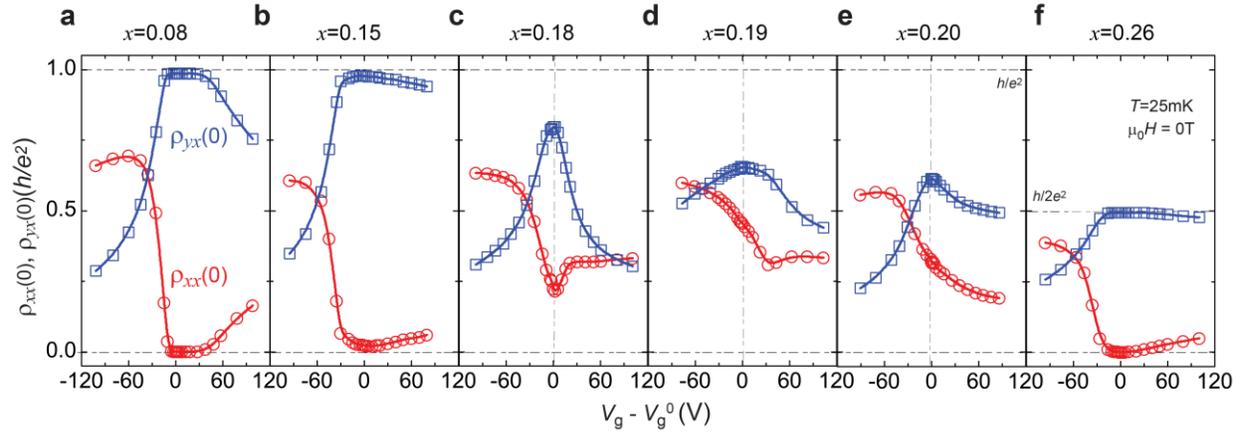

**Figure 2| Demonstration of the quantum phase transition between $C = 1$ and $C = 2$ QAH insulators. a-f**, Gate ($V_g$-$V_g^0$) dependence of $\rho_{yx}(0)$ (blue) and $\rho_{xx}(0)$ (red) of the magnetic TI/TI penta-layer heterostructures with different Cr doping $x$ in the middle magnetic TI layer. $x=0.08$ (a), $x=0.15$ (b), $x=0.18$ (c), $x=0.19$ (d), $x=0.20$ (e), and $x=0.26$ (f). All measurements were taken at $T = 25$ mK and $\mu_0 H = 0$ T after magnetic training.

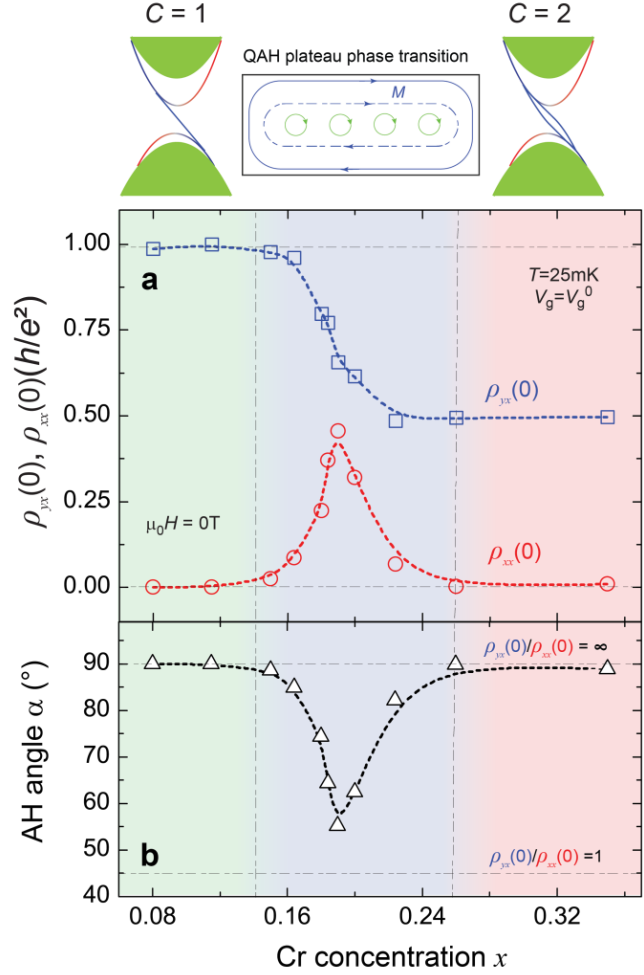

**Figure 3| Evolving chiral edge transport during the phase transition between $C = 1$ to $C = 2$ QAH insulators. a,** The Cr doping concentration $x$ dependence of $\rho_{yx}(0)$ (blue) and $\rho_{xx}(0)$ (red) of the magnetic TI/TI penta-layer heterostructures measured at $V_g=V_g^0$ and $T= 25$ mK. Top: Schematic of the quantum phase transition between the QAH insulators with one and two chiral edge channels. **b,** The Cr doping concentration $x$ dependence of the corresponding AH angle α (i.e. α = arctan $(\rho_{yx}(0)/\rho_{xx}(0))$ of the magnetic TI/TI penta-layer heterostructures. In the Chern number change-induced quantum phase transition region, the AH angle α is always greater than 45°, indicating the existence of the chiral edge transport.

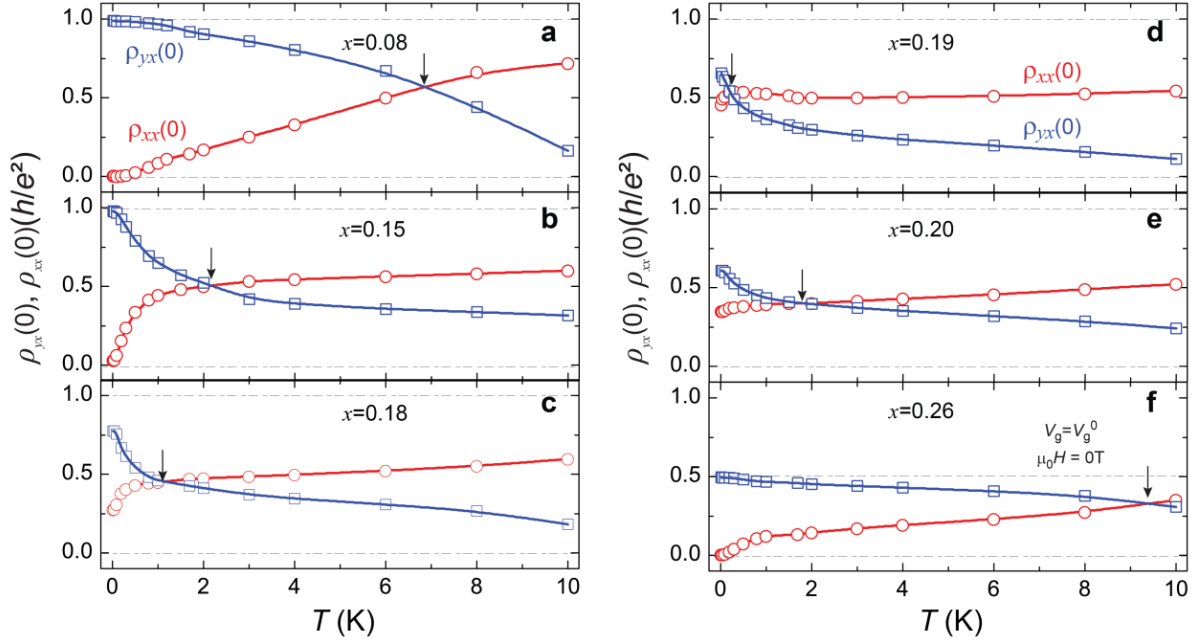

**Figure 4| Evolution of the critical temperatures of the QAH insulators. a-f,** Temperature dependence of $\rho_{yx}(0)$ (blue) and $\rho_{xx}(0)$ (red) of the magnetic TI/TI penta-layer samples with $x = 0.08$ (a), $x = 0.15$ (b), $x = 0.18$ (c), $x = 0.19$ (d), $x = 0.20$ (e), and $x = 0.26$ (f). All data were acquired at $\mu_0 H = 0$ T after magnetic training. The critical temperature $T_c$ (indicated by arrows) values are ~6.8 K, ~2.2 K, ~1.2 K, ~0.2 K, ~1.8 K, and ~9.2 K for the $x=0.08$, $x=0.15$, $x=0.18$, $x=0.19$, $x=0.20$, and $x=0.26$ samples, respectively.